\documentclass[aps,psfig,prd,preprintnumbers,amsmath,amssymb,superscriptaddress]{revtex4}
\newcommand{\vect}{\mathbf}
\usepackage{graphics}
\begin{document}
\title{Cosmological Perturbations During Radion Stabilization}
\author{P.R.~Ashcroft}
\affiliation{Department of Applied Mathematics and Theoretical
Physics, Centre for Mathematical Sciences, University of Cambridge,
Wilberforce Road , Cambridge CB3 OWA, UK}
\author{C. van de Bruck} 
\affiliation{Department of Applied Mathematics, Astro--Particle Theory \& Cosmology 
Group, Hounsfield Road, Hicks Building, University of Sheffield, Sheffield S3 7RH, 
United Kingdom}
\author{A.-C. Davis}
\affiliation{Department of Applied Mathematics and Theoretical
Physics, Centre for Mathematical Sciences, University of Cambridge,
Wilberforce Road , Cambridge CB3 OWA, UK}
\date{24 August 2004}
\begin{abstract}
We consider the evolution of cosmological perturbations during radion 
stabilization, which we assume to happen after a period of inflation 
in the early universe. Concentrating on the Randall-Sundrum brane 
world scenario, we find that if matter is present both on the positive and 
negative tension branes, the coupling of the radion to matter fields could have 
significant impact on the evolution of the curvature perturbation and 
on the production of entropy perturbations. We investigate both the case of 
a long-lived and short-lived radion and outline similarities and differences 
to the curvaton scenario.
\end{abstract}

\maketitle 

\vspace{0.5cm}
\noindent DAMTP-2004-59

\section{Introduction}
Present and future cosmological observations enable us to constrain or rule out 
models of the early universe. For example, the observations made by the Wilkinson Microwave Anisotropy 
Probe (WMAP) already put some constraints on inflationary cosmology, 
\cite{peiris,kolb,liddle,schwarz}. 
The observations are consistent with purely adiabatic perturbations with a scale 
invariant power spectrum, but some ``anomalies'', such as the small power of the 
quadrapole and features in the CMB anisotropy spectrum, have been reported, \cite{wmapresults}. 

Future cosmological observations however, will not only constrain current popular models 
of the early universe, they will also test models ``beyond the standard lore'', such as 
models based on string theory, extra dimensions, etc. For example, 
alternative mechanisms to inflation have been proposed recently. In contrast to the standard 
inflationary picture, where density perturbations are generated during inflation and then
stretched onto superhorizon scales, fields other than the inflaton 
field generate the initial perturbations in the new mechanism. In the curvaton scenario 
initial isocurvature perturbations are transformed into curvature perturbations 
by a subsequent decay of a second field. This field, dubbed the curvaton, is already present in the early universe but is dynamically unimportant during inflation, \cite{curvatoninitial,curva1,curvatonfinal,
curvatonappli1,curvatonappli2,curvatonappli3,curvatonappli4,curvatonappli5}. 
Another example is the idea of ``modulated perturbations'', \cite{modinitial,mod1,mod2,mod3,modfinal,
modappli1}. 
According to this idea, coupling constants 
(and other physical properties such as masses of the particles) are functions of 
the vacuum expectation value of fundamental 
(light) scalar fields present during inflation. Because of vacuum fluctuations 
in these fields, different regions in spacetime have different values of coupling 
constants. These fluctuations can be converted into curvature perturbations, for 
example during reheating which is not homogeneous in space if the decay rate itself 
fluctuates. Note that both the curvaton scenario as well as the idea of modulated 
perturbations need additional scalar fields in the theory, which are dynamically unimportant-- at least initially, during inflation. 

In this paper we do not seek an alternative to the standard picture of how 
perturbations are generated, but rather investigate how far the stabilization 
of moduli fields can alter the evolution of cosmological perturbations generated 
during a period of inflation in the early universe. Moduli fields appear in theories 
beyond the standard model based on supersymmetry or superstring theory. In some models
they can be long-lived and the life-time can be even larger than the age of the universe. 
The prime example of a moduli field is the radion, measuring the size of the extra 
dimensional spacetime in brane world scenarios (see e.g. \cite{branerev1,branerev2,branerev3} 
for recent reviews on brane cosmology). One important 
property of the radion is that it couples explicitly to matter fields on the branes 
(see e.g. \cite{Brax:2002nt} and references therein). 
Depending on the details of the bulk geometry and the matter content on the branes, 
the coupling is field dependent and can become quite large. It is the coupling between 
the different matter forms to the radion which can lead to interesting consequences for 
the evolution of perturbations. Furthermore, the radion can be short or long-lived 
depending on the stabilization mechanism. In the case of the Goldberger-Wise stabilization 
mechanism in the Randall-Sundrum brane world, \cite{goldbergerwise}, the radion 
obtains a mass of the order of TeV, which makes is short-lived, \cite{csaki}. In this paper we 
investigate the impact of radion stabilization 
on the cosmology and dynamics of cosmological perturbations. We consider both 
the case of a short-lived and long-lived radion; in the latter, the radion will 
constitute some (or all) of the dark matter in the universe. Throughout the paper
we work in the Randall-Sundrum scenario, \cite{RS}, because the coupling of the radion to matter
on both branes can be quite large and new effects will appear (see e.g. \cite{cmb1,cmb2} for the 
effects of the radion on the CMB anisotropies). In \cite{periradion,kolbradion,perivernon} 
the radion and its r{\^o}le in cosmology were discussed in considerable depth. Here we take 
into account the warped geometry of the bulk.

The paper is organized as follows: in Section~\ref{sect:eom} we will present and describe the 
low-energy effective theory of the Randall-Sundrum brane world. We then present
the equations of motion for both background and perturbations. 
In Section~\ref{sect:radevo} we will discuss the evolution of the radion and the different 
constraints on the theory coming from  nucleosynthesis,  overclosure of 
the universe and the amount of entropy perturbations generated. In 
Section~\ref{sect:radstab} we present our numerical results of the evolution of the curvature 
and entropy perturbations. Our conclusions are presented in Section~\ref{sect:radconc}.

\section{The Effective Action and Equations of Motion}\label{sect:eom}
The form of the low-energy action for the Randall-Sundrum brane world scenario, in
which two boundary branes are embedded in a slice of an Anti-de Sitter (AdS) spacetime, 
has been studied in the literature already, see e.g. 
\cite{kannosoda,chiba,Brax:2002nt,palmadavis1,palmadavis2} 
and references therein. At low energies the theory is a scalar-tensor theory with 
specific matter couplings. In the Einstein frame it takes the form
\begin{eqnarray}
\mathcal{S}_{\rm EF} &=&  \int
d^4x \sqrt{-g} \left[ \frac{1}{16\pi G} \mathcal{R} - \frac{1}{2} (\partial R)^2 - V(R) \right] \nonumber \\
&& \hspace{1cm} + \mathcal{S}_m^{(1)} ( \Psi_1, A^2(R) g_{\mu\nu} ) 
+ \mathcal{S}_m^{(2)}
( \Psi_2, B^2(R) g_{\mu\nu} ).
\end{eqnarray}
We have allowed for the possibility of two matter forms, each of which is confined to the positive or negative tension 
brane. The couplings between the fields on the branes and the field $R$ are described by the 
functions $A(R)$ and $B(R)$. The coupling to matter has some interesting consequences, 
the primary feature being that we no longer have energy-momentum conservation for each component. 
In fact
\begin{eqnarray}\label{emconservation}
\nabla_\mu T^{\mu\nu}_i =  \alpha_R^i ( \partial^\nu R) T_i,
\end{eqnarray}
where $i$ describes the particular brane in hand. Note that the field couples to 
the trace of the energy-momentum tensor and so there is no coupling
to radiation. One defines the coupling functions 
\begin{eqnarray}
\alpha_R^{(1)} = \frac{\partial \ln A}{\partial R}, \ \ \ \ \alpha_R^{(2)} 
= \frac{\partial \ln B}{\partial R}.
\end{eqnarray}
The energy conservation equation~(\ref{emconservation}) is modified 
in this theory because the masses of particles on 
the branes are functions of the radion, and hence can vary with time if the 
radion varies with time. Alternatively, we could choose a frame in which the 
masses of the particles on the positive tension brane are constant. In this
frame, the masses of the particles on the negative tension brane as well as 
the four-dimensional Planck mass vary with time. We will study the theory in the 
Einstein frame, because after stabilization of the radion the frames agree.

One problem common to nearly all models with moduli fields is how to stabilize these fields. 
We shall include a potential of the form 
\begin{eqnarray}
V(R) = \frac{1}{2} M_R^2 ( R - R_c )^2,
\end{eqnarray}
but here we note that the origin of such a potential might have to be derived from non-perturbative
effects in the underlying theory. As we will see below, the mass $M_R$ is constrained 
by  requiring that the field does not overclose the universe and also by nucleosynthesis.  

Finally, the couplings $A(R)$ and $B(R)$ have the form 
\begin{eqnarray}
A(R) &=& \cosh\left(\frac{R}{\sqrt{6}\, M_{pl}}\right), \ \ \ \alpha_R^{(1)} = 
\frac{1}{\sqrt{6}\, M_{pl}}\tanh\left(\frac{R}{\sqrt{6}\, M_{pl}}\right), \label{eq:ar}\\
B(R) &=& \sinh\left(\frac{R}{\sqrt{6}\, M_{pl}}\right), \ \ \ \alpha_R^{(2)} = 
\frac{1}{\sqrt{6}\,M_{pl}}\mathrm{coth}\left(\frac{R}{\sqrt{6}\, M_{pl}}\right).\label{eq:br}
\end{eqnarray}
The geometry of the higher-dimensional spacetime is encoded in these coupling
functions. Depending on the value of $R$, the couplings can be quite large. For small values of $R$, 
the function $\alpha^{(2)}_R$ behaves as $\alpha^{(2)}_R \sim 1/R$, whereas one finds 
 $\alpha^{(1)}_R \sim R / M_{pl}^2$. 

The theory described above is a good description of the two-brane system as long as the energy
densities on both branes are much less than the brane tension. All length scales considered should 
be much larger than the curvature scale of the Anti-de Sitter bulk. Furthermore, we have assumed 
that Kaluza-Klein modes are irrelevant at the energy scales we consider. 

Primarily we are concerned with the application to cosmology. We allow matter to be 
present on both branes, each with an energy density $\rho_m^{(1)}$ and $\rho_m^{(2)}$
respectively. We also include radiation with energy density $\rho_\gamma$. To end up with 
a realistic cosmology, i.e. with late time acceleration, we include a cosmological constant 
$\Lambda$. Assuming a flat universe, the equations for the background read:
\begin{eqnarray}
H^2 &=& \frac{1}{3 M_{pl}^2} \left( \frac{1}{2} \dot{R}^2 +
U\left(  R \right) + \rho_m^{(1)} + \rho_m^{(2)}
+ \rho_\gamma + \rho_\Lambda\right),\\
\ddot{R} + (3H + \Gamma ) \dot{R} &=& - \left[ \frac{\partial U}{\partial R} +
  \alpha_R^{(1)} \rho_m^{(1)} +   \alpha_R^{(2)} \rho_m^{(2)}  \right],\label{eq:rdot}\\
\dot{\rho}_m^{(i)} &=& - 3H \rho_m^{(i)} +  \alpha_R^{(i)}
\dot{R} \rho_m^{(i)},\ \ \ j = 1,2, \\
\dot{\rho}_\gamma &=& - 4H \rho_\gamma + \Gamma \dot{R}^2.
\end{eqnarray}
We will define the density parameter for the different components by 
$\Omega_i = \rho_i/\rho_{cr}$ as is usual.

The equations of motion of perturbations in matter, the radion field 
and the metric can be obtained from the effective action. 
We consider scalar perturbations only in this paper and 
work in the longitudinal gauge, in which the perturbed metric has the form 
(neglecting any anisotropic stress),
\begin{eqnarray}
ds^2 = - ( 1 + 2\Psi ) dt^2 + a^2(t) ( 1 - 2 \Psi ) d\vect{x}^2.
\end{eqnarray}
The energy-momentum tensor for the two fluids has the form
\begin{eqnarray}
{T^{\mu}}_\nu = \left( \begin{array}{cc} - ( \rho + \delta\rho ) &
  -\frac{1}{a^2} \partial^i \delta q \\
\partial_j \delta q & ( p + \delta p ) {\delta^i}_j
  \end{array}\right).
\end{eqnarray}
According to our assumptions, $\delta p_i = \omega_i \delta \rho_i$ for both matter and 
radiation. Then, expanding in Fourier modes, the perturbation equations read at first order
\begin{eqnarray}
\delta \ddot{R} + ( 3H + \Gamma ) \delta \dot{R} + \frac{k^2}{a^2} \delta
R + \frac{\partial^2 V}{\partial R^2}\delta R
 &=&  - \alpha_R^{(1)} 
\delta \rho_m^{(1)} - \alpha_R^{(2)} 
\delta \rho_m^{(2)} - \left( \frac{\partial\alpha_R^{(1)}}{\partial R} 
\rho_m^{(1)} + \frac{\partial \alpha_R^{(2)}}{\partial R} 
\rho_m^{(2)} \right) \delta R \nonumber \\ 
&& - 2 \Psi \left( \frac{\partial V}{\partial R} +
\alpha_R^{(1)} \rho_m^{(1)} + \alpha_R^{(1)} \rho_m^{(1)} + \Gamma
\dot{R} \right) + 4 \dot{\Psi}\dot{R}, \label{eq:pert1}\\
\delta \dot{\rho}_m^{(j)} + 3 H \delta \rho_m^{(j)} - 3 \dot{\Psi} \rho_m^{(j)} +
\frac{k^2}{a^2} \delta q_m^{(j)} &=& Q_m^{(j)}\Psi + \delta Q^{(j)}_m, \ \ j = 1,2 \\
\delta \dot{\rho}_\gamma + 4 H \delta \rho_\gamma - 4 \dot{\Psi} \rho_\gamma +
\frac{k^2}{a^2} \delta q_\gamma &=& Q_\gamma \Psi + \delta Q_\gamma,\\
\delta\dot{q}_m^{(j)} + 3 H \delta q_m^{(j)} + \Psi \rho_m^{(j)} &=& 
-\alpha_R \rho_m^{(j)} \, \delta R,\\
\delta\dot{q}_\gamma + 3 H \delta q_\gamma + \frac{4}{3} \rho_\gamma \Psi +
\frac{1}{3} \delta \rho_\gamma &=& 0,\\
3 H ( \dot{\Psi} + H \Psi ) - \frac{k^2}{a^2} \Psi&=& - \frac{1}{2M_{pl}^2} \delta \rho.\label{eq:pert2}
\end{eqnarray}
where we have used the abbreviations
\begin{eqnarray}
Q_m^{(j)} &=&  \alpha_R^{(j)} \dot{R}  \rho_m^{(j)},\label{eq:Q1} \\
\delta Q^{(j)}_m &=& \left( \alpha_R^{(j)} \delta
\dot{R} +   \frac{\partial \alpha_R^{(j)}}{\partial R} \dot{R} \delta R\right) \rho_m^{(j)}
+  \alpha_R^{(j)} \dot{R}  \delta
\rho_m^{(j)} - \Psi Q_m^{(j)},  \ \ \ j = 1,2. \\ 
Q_\gamma &=& \Gamma \dot{R}^2,\\ 
\delta Q_\gamma &=&  2\Gamma \dot{R} \delta\dot{R} - \Psi \Gamma  \dot{R}^2,\label{eq:Q2}\\
\delta \rho &=& \delta \rho_\gamma + \delta \rho_{m}^{(1)} + \delta \rho_{m}^{(2)} 
+ \dot{R}\delta \dot R + \frac{\partial V}{\partial R}\delta R - \Psi\dot{R}^2.
\end{eqnarray}

The fluids and the radion field exchange energy during the cosmological expansion due
to the non-vanishing coupling. This is to be compared with the original curvaton scenario, 
in which the curvaton field decays into radiation (and matter). 

Furthermore, there is also energy transfer between the radion field 
and the radiation fluid due to the $\Gamma$ interaction term. The decay rate is given as, \cite{mazumdar}, 
\begin{eqnarray}
\Gamma = a \frac{M_R^3}{192 \pi M_{pl}^2},
\end{eqnarray}
where the constant $a$ is determined by the number of extra dimensions
and the compactification and is ${\cal O}(1)$. We shall set $a = 1$
for the duration of this paper. This means the radion
has a decay time of
\begin{eqnarray}
\tau_R \approx 1.2 \times 10^7 \mathrm{yr} \
\left(\frac{\mathrm{GeV}}{M_R} \right)^3.
\end{eqnarray}
Thus for $M_R < 0.1 \, {\rm GeV}$, the radion
would effectively be stable over the lifetime of the 
universe. If we require the radion to decay before nucleosynthesis,
then we need 
\begin{eqnarray}
\Gamma \geq H( t_{\mathrm{nucl}} ) \sim 10^{-43}\,  M_{pl}.
\end{eqnarray}
This would then require a radion mass $M_R \gtrsim 5 \times 10^{-14}\, 
M_{pl} \approx  10^5 \, \mathrm{GeV}$, assuming the constant $a$ is of order unity. 
The radion stabilizes at  $ H \sim M_R$ so this gives a long period of radion domination. 

The evolution of the perturbation quantities is described in full by
equations (\ref{eq:pert1}-\ref{eq:pert2}). However, it will be helpful
for our understanding to recast these. In general, the density perturbations, $\delta \rho_i$, and the metric
perturbation, $\Psi$, are gauge dependent. It is more common to work with 
gauge-invariant quantities. One important quantity, the curvature perturbation on constant density hypersurfaces,
 is (see e.g. \cite{bardeen,brandenbergerreview,malik1})
\begin{eqnarray}
\zeta = \Psi + H \frac{\delta \rho}{\dot{\rho}}.
\end{eqnarray}
Note that this differs in sign from some of the definitions in the literature.
In addition we can define the curvature perturbation on uniform
$i$-fluid density hypersurfaces as, \cite{malik1},
\begin{eqnarray}
\zeta_i = \Psi + H \frac{\delta \rho_i}{\dot{\rho}_i}.
\end{eqnarray}
With this prescription, it is clear that the total curvature
perturbation is a weighted sum of the individual components,
\begin{eqnarray}
\zeta = \sum_i \frac{\dot{\rho_i}}{\dot{\rho}} \zeta_i.
\end{eqnarray}
Furthermore, we are able to define a relative entropy perturbation
between two components as 
\begin{eqnarray}
\mathcal{S}_{ij} = 3 ( \zeta_i - \zeta_j ).
\end{eqnarray}

To make some predictions about the behavior of the perturbations we
work in the ``separate universes'' picture, which effectively allows us to
ignore the spatial gradient terms. In the long-wavelength limit, the
perturbed continuity equation becomes
\begin{eqnarray}
\delta\dot{\rho} + 3H ( \delta \rho + \delta P ) = 3 ( \rho + P ) \dot{\Psi}, \label{eq:pert3}
\end{eqnarray}
where we define
\begin{eqnarray}
&&\rho = \sum_i \rho_i, \ \ P = \sum_i P_i, \ \ \delta \rho = \sum_i
\delta \rho_i, \ \ \delta P = \sum_i \delta P_i.
\end{eqnarray}
If one rewrites equation (\ref{eq:pert3}) in terms of the total
curvature perturbation, $\zeta$, it is relatively straightforward to show
\begin{eqnarray}
\dot{\zeta} = \frac{H}{\rho + P} \, \delta P_{\mathrm{nad}},\label{eq:zetaevo}
\end{eqnarray}
where the non-adiabatic pressure perturbation is $\delta
P_{\mathrm{nad}} \equiv \delta P - c_s^2 \delta\rho$ and the adiabatic
sound speed is $c_s^2 = \dot{P}/\dot{\rho}$. This means that the total
curvature perturbation will be constant on large scales for purely
adiabatic perturbations. In the context of brane worlds, where energy 
can be exchanged with the bulk space time, this quantity is not necessarily 
constant (see e.g. \cite{zeta1,zeta2,zeta3}).

It is possible to break this down further so that one may write, \cite{curvatonappli2},
\begin{eqnarray}
\delta P_{\mathrm{nad}} &\equiv& \delta P_{\mathrm{int}} + \delta
 P_{\mathrm{rel}},\\
&=&\sum_i ( \delta P_i - c_i^2 \delta \rho_i) + \frac{1}{6H\dot{\rho}}
 \sum_{i,j} \dot{\rho_i}\dot{\rho_j} ( c_i^2 - c_j^2) \mathcal{S}_{ij},\label{eq:deltaPnad}
\end{eqnarray}
where $c_i^2 \equiv \dot{P_i}/\dot{\rho_i}$ is the adiabatic sound speed of
the fluid and is related to $c_s$ by
\begin{eqnarray}
c_s^2 = \sum_i \frac{\dot{\rho}_i}{\dot{\rho}} c_i^2.
\end{eqnarray}
Further, let us define
\begin{eqnarray}
\delta P_{{\rm int},i} \equiv \delta P_i -c_i^2 \delta \rho_i,
\end{eqnarray}
so that $\delta P_{{\rm int}} = \sum_i \delta P_{\mathrm{int},i}$.

On superhorizon scales it is possible to write the individual perturbation equations 
in the form, \cite{curvatonappli2},
\begin{eqnarray}
\dot{\zeta_i} = - \frac{3H^2}{\dot{\rho}_i} \left( \delta P_i - c_i^2
\delta \rho_i \right) + \frac{H Q_i}{\dot{\rho}_i} \left[ \frac{\delta
    Q_i}{Q_i} + \left( \frac{\dot{\rho}}{2\rho} -
    \frac{\dot{Q}_i}{Q_i} \right)\frac{\delta\rho_i}{\rho_i} +
    \frac{\dot{\Psi}}{H} + \Psi \right].\label{eq:zetaievo}
\end{eqnarray}
The $Q_i$ and $\delta Q_i$ are given in equations
(\ref{eq:Q1}-\ref{eq:Q2}) with $Q_R = - Q_m^{(1)} - Q_m^{(2)} -
Q_\gamma$ and similarly for $\delta Q_R$. From this, one can immediately see 
that for non-interacting perfect fluids with $Q_i = 0$ and 
$\delta P_i = c_i^2 \delta \rho_i$, the individual curvature 
perturbations for each fluid remain constant on large scales.

There is now large scope for the setting of the initial conditions of
the perturbations. We shall assume that the initial conditions result from  a period of inflation in the early universe. Then, it is known 
that any light scalar field, $\phi_j$, will pick up a perturbation, \cite{lyth,riotto},
\begin{eqnarray}
\delta \phi_j = \left(\frac{H}{2\pi}\right)_*, \ \ \ \ \  \delta\dot{\phi}_j = 0,
\end{eqnarray}
where this is evaluated at horizon crossing for the inflaton
field. This depends on the scale of inflation but if this is to be
taken as $V \lesssim 1.5 \times 10^{-8} \, M_{pl}^4$, \cite{liddle}, this tells us
that
\begin{eqnarray}
|\delta R| \lesssim 1.1 \times 10^{-5}\,  M_{pl}.
\end{eqnarray}
 If this results in too much entropy production, this
will offer a constraint on the scale of inflation.
The remaining initial conditions are arranged to be adiabatic initially,
as given in \cite{ma},
\begin{eqnarray}
\Psi = \frac{4}{3}C, \ \ \ \delta \rho_m^{(j)} &=& - 2 C
     \rho_m^{(j)}, \ \ \ \delta \rho_\gamma = - \frac{8}{3} C
     {\rho_\gamma}, \label{eq:pertic}
\end{eqnarray}
where $C$ is some initial scale to be decided in advance. We shall set
$C = 10^{-5}$ in accordance with current observations. Since we are
dealing with linear perturbation theory, it is only the ratio $|\delta
R|/C$ that is important rather than their absolute values. In fact, in
future we shall normalize the curvature perturbation and plot 
$\zeta_i/ \zeta_{\mathrm{ini}}$. Note that these initial conditions are different 
from those chosen in \cite{cmb2} where $\delta R = 0$ initially. For the duration of this paper we shall set
\begin{eqnarray}
\delta R = -1.1 \times 10^{-5}\, M_{pl},
\end{eqnarray}
since this will give the largest possible effects. 

With this prescription, it is worth discussing what we might expect to
see. From equations (\ref{eq:zetaevo}) and (\ref{eq:deltaPnad}) one can
see that $\zeta$ is sourced in two ways. Firstly when $P_i \neq
P_i(\rho_i)$, which makes the first term in (\ref{eq:deltaPnad})
non-zero, and secondly if there is an entropy perturbation present. For
perfect fluids, $\delta P_{\mathrm{intr},i}$ will be zero but for a
scalar field this is not necessarily the case. Therefore we will always generate
evolution in $\zeta_R$ and thus in $\zeta$. Furthermore, our initial conditions,
whilst adiabatic between the fluid components, necessarily produce an
entropy perturbation between the radion and the fluids. This all
holds without any couplings. One should add that since the total
curvature perturbation is a weighted sum of the individual components,
the effect of the scalar field on $\zeta$ is very much dependent on
its energy density fraction. During radiation domination, $\zeta \sim
\zeta_\gamma$ which will be constant-- if $\Gamma \neq 0$ this no
longer holds. During matter domination it is then described by
\begin{eqnarray}
\zeta \approx \Omega_m^{(1)} \zeta_m^{(1)} +  \Omega_m^{(2)}
\zeta_m^{(2)} + \Omega_R \zeta_R.\label{eq:zetarho}
\end{eqnarray}
As one moves into the $\Lambda$ dominated phase this should then
become constant again. If couplings are introduced then, from
 equation (\ref{eq:zetaievo}), we know that all the $\zeta_i$ evolve. During the
energy transfer between the fields, $\delta P_{\rm{int},i} \neq 0$ and
we will also produce entropy perturbations. We should expect to see further
evolution in $\zeta$, which will be dependent on the size of the
coupling terms.

If one includes coupling between the radion and radiation only ($\Gamma
\neq 0, \alpha_R^{(j)} = 0$), we
should see an effect in $\zeta_R$, and therefore $\zeta$, as soon
as the radion stabilizes, because it becomes the dominant component. All of
its energy is transfered to radiation as it decays. Once it has
decayed, $\zeta_R$ gives no contribution to the total. However, since
there is no coupling to matter, the $\zeta_m^{(i)}$ have remained
constant and $\zeta$ will return to this value. Of course, this
conclusion will be modified if the coupling terms are non-zero.

\section{Cosmological Evolution of the Radion}\label{sect:radevo}
Before we discuss the evolution of the perturbations, we will consider the 
evolution of the radion during the inflationary epoch and the subsequent 
radiation and matter dominated epochs. We also discuss cosmological constraints 
the theory has to fulfill. We begin with the inflationary era. 
\subsection{The Epoch of Inflation}
For definitiveness we consider an inflaton field confined to the positive tension 
brane and a chaotic inflation potential of the form 
$V(\phi) = M_\phi^2 \phi^2 / 2$, where $M_\phi$ is the mass of the 
inflaton field. Further, because we want to study purely adiabatic initial conditions between
matter and radiation, we assume that during the last 60 e-folds 
the inflaton field dominated the energy density of the universe. We are not interested
in the case in which the radion was the inflaton field. With these assumptions, there 
are three cases of interest. In the first case, the mass of the radion is much 
larger than the mass of the inflaton field, i.e. $M_R \gg M_\phi$. Depending on the 
initial conditions for $R$, the radion field could have dominated the initial expansion of the early
universe. In the second case, the masses of the radion and inflaton field are 
comparable. In this instance, the energy density of the radion 
must have been much smaller than the energy density of the inflaton to comply with our assumptions. Finally, the radion can be much lighter than the inflaton field. Let us briefly discuss the 
three cases separately. 

\subsubsection{$M_R \ll M_\phi$}
In this regime, the radion is very light compared to the inflaton field. Its potential energy 
can be neglected and the radion field is driven by the coupling to the inflaton field only. 
As established earlier, \cite{ashcroft3}, 
the radion field is driven towards small values.

\subsubsection{$M_R \approx M_\phi$}
Here a period of double inflation can be realized, where each phase is driven
by either the radion or the inflaton field. In accordance with our assumptions, we  
consider the case in which the second period of inflation, driven by the inflaton field, 
is longer than 60 e-folds.  With this prescription, the radion and the inflaton field stabilize at the same time.



\subsubsection{$M_R \gg M_\phi$}

This case is very similar to the second example, except that the radion now stabilizes  
much earlier. Initially, it is possible that the radion dominated the expansion rate of the universe 
before settling into the minimum of the effective potential. In principle, the radion 
could have set up the initial conditions for a period of inflation.



In the following, we consider a scenario where the radion decays well after inflation, 
which corresponds to the first case only. This ensures that the perturbations 
created during inflation are adiabatic as is the case in single field inflation.

\subsection{The radiation and matter dominated epochs}
The behavior of the radion field depends on its mass. 
If it is very light, the lifetime can be very large and in fact, in this instance, 
we assume that the lifetime is much larger than the age of the universe. 
However, depending on the origin of the potential for the radion, it could 
also be very heavy. In the case of radion stabilization via the Goldberger-Wise 
mechanism the radion mass is of order TeV. Here, the radion decays 
into radiation and matter in the very early universe. We will consider these 
two cases in the rest of the paper. Further, we need to make assumptions about
where the matter fields are located. In principle, both the standard model 
particles as well as the dark matter particles can be located only on the 
positive tension brane, whereas there is no matter on the negative tension 
brane. Alternatively, on can imagine the situation where only the standard
model particles are confined on the positive tension brane, whereas the 
dark matter particles are confined on the negative tension brane. 
Although mixtures of these scenarios can be imagined, in this paper we will allow only 
baryons to live on the positive tension brane; the dark matter being made up from matter on the negative tension brane
and from the oscillating radion field.

In order to make our study compatible with present day observation, we
take values for the parameters as given by the most recent WMAP data, \cite{wmapresults}. 
This means that today our universe should be almost flat and made up of
\begin{eqnarray}
\Omega_\Lambda = 0.73, \ \ \ \Omega_{\mathrm{DM}} = 0.23, \ \ \
\Omega_M = 0.04.
\end{eqnarray}
Furthermore, we ensure that matter-radiation equality occurs in the
range obtained by the WMAP team, 
\begin{eqnarray}
3200 \lesssim z_{eq} \lesssim 3400.
\end{eqnarray}

Nucleosynthesis makes excellent predictions for the abundances of the light 
elements and we require that these predictions are not modified too much. 
Nucleosynthesis gives a constraint on the expansion rate $H_{\rm nucl}$ 
at the time when the light elements form. If the particle masses change during 
the history of the universe, as is the case in our theory, this will 
influence the time dependence of $H$ and therefore the expansion rate at 
the time of nucleosynthesis. The evolution of the masses is specified by 
the functions $A$ and $B$. The requirement that any changes of $H_{\rm nucl}$ 
are constrained by nucleosynthesis gives, 
\cite{bartolo},
\begin{equation}
\frac{A_{nucl} - A_0}{A_0} \leq 0.1,
\end{equation}
where $A_{nucl}$ and $A_0$ are the coupling functions at the time of nucleosynthesis 
and today respectively. A similar bound is valid for $B$ if some matter form is present 
on the negative tension brane. Note that these considerations neglect variations in the 
reaction rates,  which itself are functions of masses-- and therefore 
of $A$-- if the standard model particles live on the positive tension brane. In contrast, if 
they live on the negative tension brane instead, the reaction rates would 
be functions of $B$. Therefore, the constraint above is a rather conservative one.

If the radion is light ($M_R < 0.1 \, {\rm GeV}$), we have seen that it
is effectively stable over the lifetime of the universe. It can be
shown, \cite{turner}, that for a scalar field, $\psi$, oscillating in the
minimum of its potential, $V \sim \psi^n$, the energy density scales 
\begin{eqnarray}
\rho_\psi \sim a^{-6n/(n+2)}.
\end{eqnarray}
In our case this would mean the radion, $R$, behaving like
non-relativistic pressureless matter. However, we also have the extra
source terms, $\alpha^{(j)} \rho_m^{(j)}$, on the right hand side of
equation (\ref{eq:rdot}) that could amend this result. The
oscillations of $\dot{R}$ are damped by the Hubble parameter and the
energy density $\rho_m^{(j)}$ will scale approximately like
$a^{-3}$. This means that the source terms will decay faster than the
others in equation~(\ref{eq:rdot}). Therefore, we should still find that the
radion behaves like matter when it oscillates in the minimum of its
potential, regardless of the new couplings. This is confirmed by our
numerical results. In fact, the couplings are most important for
the background solutions as the radion rolls down the potential. In
practice, this transfers energy between the different species. 

Since $R$ behaves like matter when oscillating at its minimum, 
this allows us to investigate a number of possibilities:
\begin{enumerate}
\item $R$ plays the r\^ole of dark matter alone.
\item $R$ is part of the dark matter with the remaining contribution
  from the positive and/or negative tension brane.
\item $R$ is subdominant with the matter on the second brane making up
  all of the dark matter.
\end{enumerate}

Stabilization occurs when $H \sim M_R$ and from this point
$\rho_R \sim a^{-3}$ (which inevitably succeeds $\rho_\gamma \sim a^{-4}$). 
Therefore, one can not stabilize the radion too early as this would lead to premature 
matter domination. This means that the radion cannot be
too heavy. Moreover, if the radion is very light, then the background
dynamics are more sensitive to the couplings. This may result in too
large an evolution of $R$. Some of this can be compensated for in the
initial conditions, but in general we shall find it difficult to
generate extremes of masses. If $R$ is to play the r\^{o}le of dark
matter, then its mass should be around $M_R \sim 10^{-50} \,M_{pl}$,
\cite{arbey1,arbey2}, to agree with galactic halo results. We
shall take this to be given in this instance and match our initial
conditions to the correct cosmology. The situation in which $R$ is
always subdominant may be interesting because even though we can not
see the field, its coupling to matter may have some interesting
effects on the perturbations. 

Whilst it is possible to have dark matter-- in addition to baryons-- present 
on the positive tension brane, the coupling function in equation~(\ref{eq:ar}) 
is small in general. In contrast, it is possible to produce  large couplings 
on the negative tension brane, as it can be seen from equation~(\ref{eq:br}).

\section{Radion Stabilization}\label{sect:radstab}
Having studied the background evolution, we now turn our attention to the 
evolution of the perturbations. We first discuss the case of a long-lived, i.e. 
stable radion field, before we consider the effects of radion decay into radiation.

\subsubsection{The case of a long-lived radion}
\begin{figure}[!ht]
\centerline{\scalebox{0.7}{\includegraphics{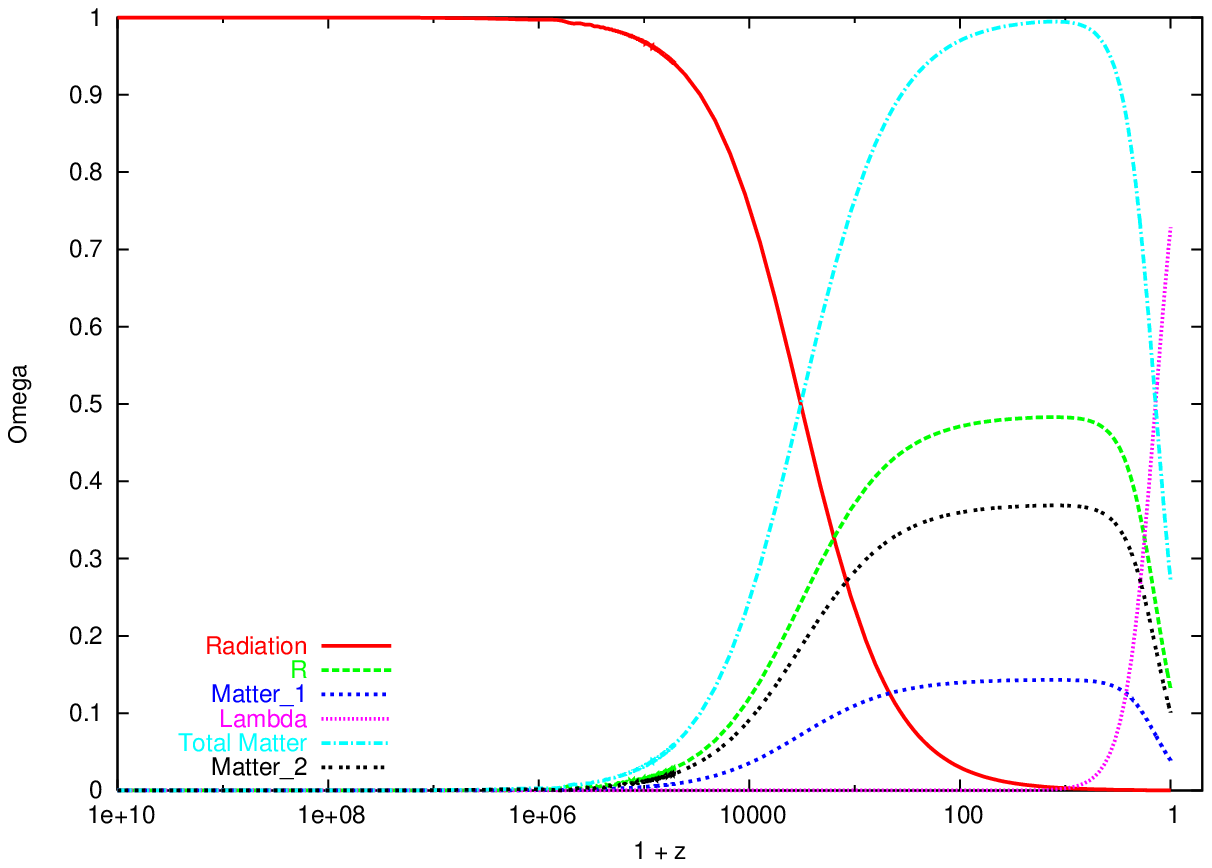}\includegraphics{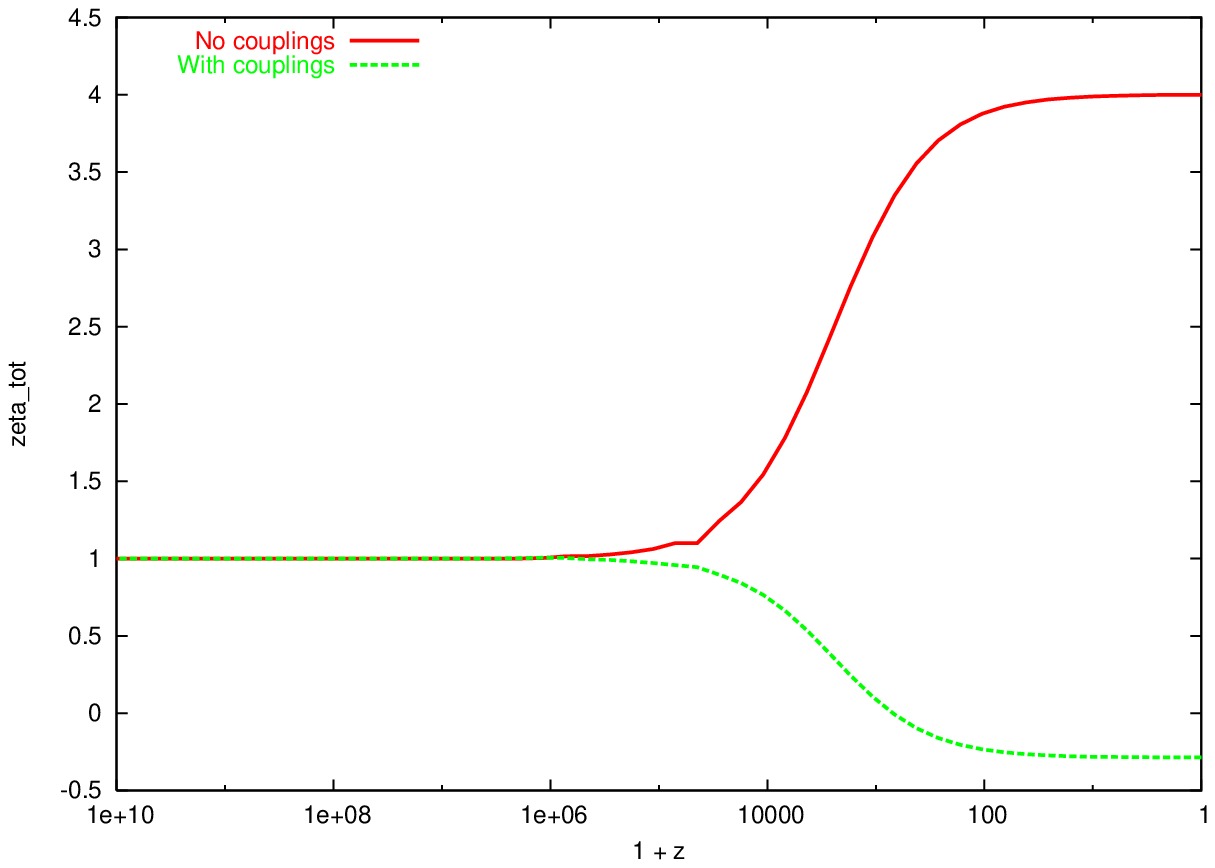}}}
\centerline{\scalebox{0.7}{\includegraphics{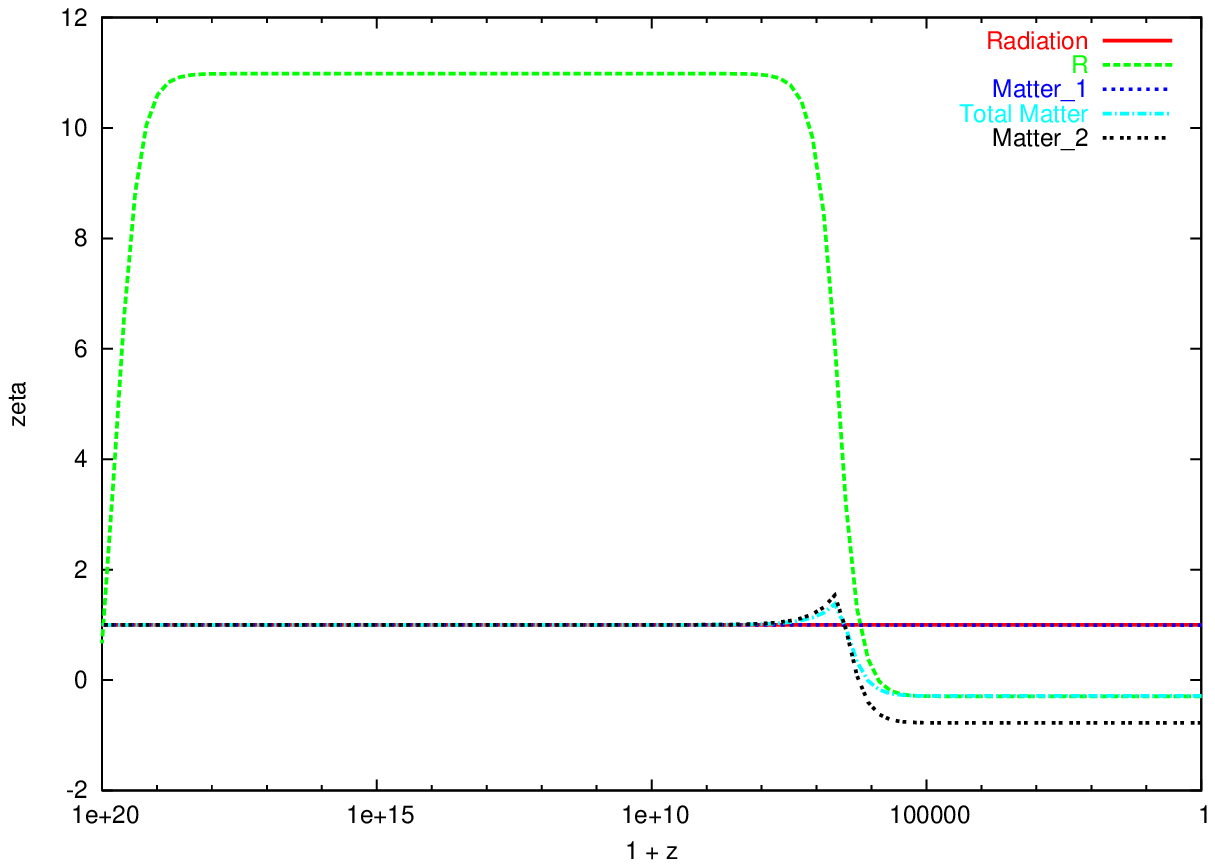}\includegraphics{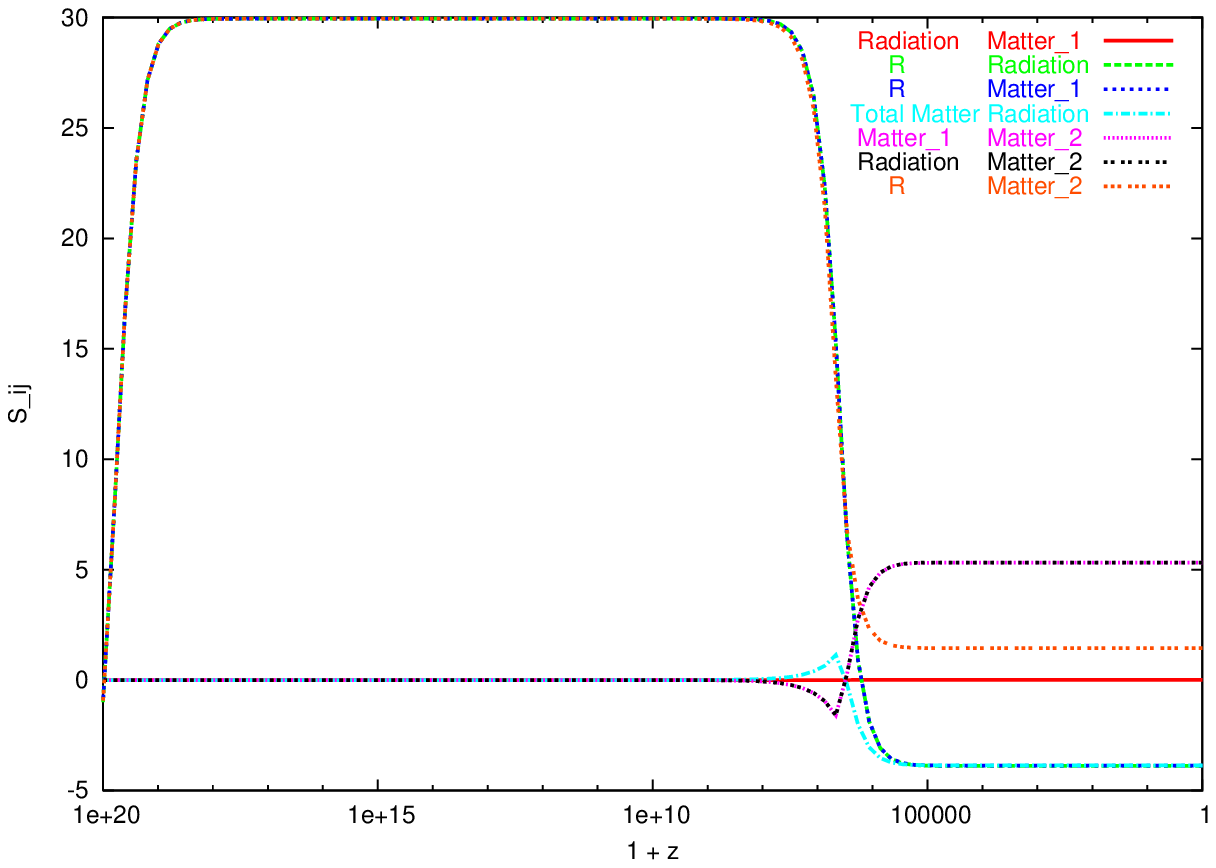}}}
\caption{This plot shows the results for a universe with $\Omega_\Lambda = 0.73$,
  $\Omega_R = 0.13$, $\Omega_{m}^{(2)} = 0.10$, $\Omega_{m}^{(1)} =
  0.04$. Furthermore, we set $M_R = \sqrt{6}\times 10^{-50}\,M_{pl}$ . In the top left we show
  the evolution of the density parameters, $\Omega_i$. In the top right plot we see the 
  evolution of the total curvature perturbation, $\zeta$, which demonstrates the effect of 
  the couplings $\alpha_R^{(j)}$, compared with the uncoupled case.  We set $R_0
  = 0.10\,M_{pl}$ and $R_c = 0.03\,M_{pl}$ so that $10^{-4} \lesssim \alpha_R^{(1)}\,M_{pl} \lesssim 10^{-2}$ and
  $10 \lesssim \alpha_R^{(2)}\,M_{pl} \lesssim 10^3$. The
  lower left plot shows the behavior of all the individual $\zeta_i$
  and on the right we show the evolution of all entropy perturbations, $S_{ij}$, between 
  the different species.}\label{fig:msa2}
\end{figure}

In the Randall-Sundrum model, the coupling of the radion to matter on the positive tension brane 
is never large, $\alpha_R^{(1)} \leq 1/\sqrt{6}$. In contrast, the 
coupling to matter on the negative tension brane has
the property $\alpha_R^{(2)} \sim 1/R$ for small $R$. This allows the
possibility of generating rather large couplings. In our first
example, we take the radion to be approximately half of the dark
matter. The results are shown in Figure~\ref{fig:msa2}. Here we are able
to generate couplings as large as $\alpha_R^{(2)} \sim 1000$ and unsurprisingly, this gives
some strong effects on $\zeta$. One sees that there are large entropy
perturbations and some interesting changes in the individual
$\zeta_i$. The most striking feature is that $\zeta$ evolves to negative values. 

\begin{figure}[!ht]
\centerline{\scalebox{0.7}{\includegraphics{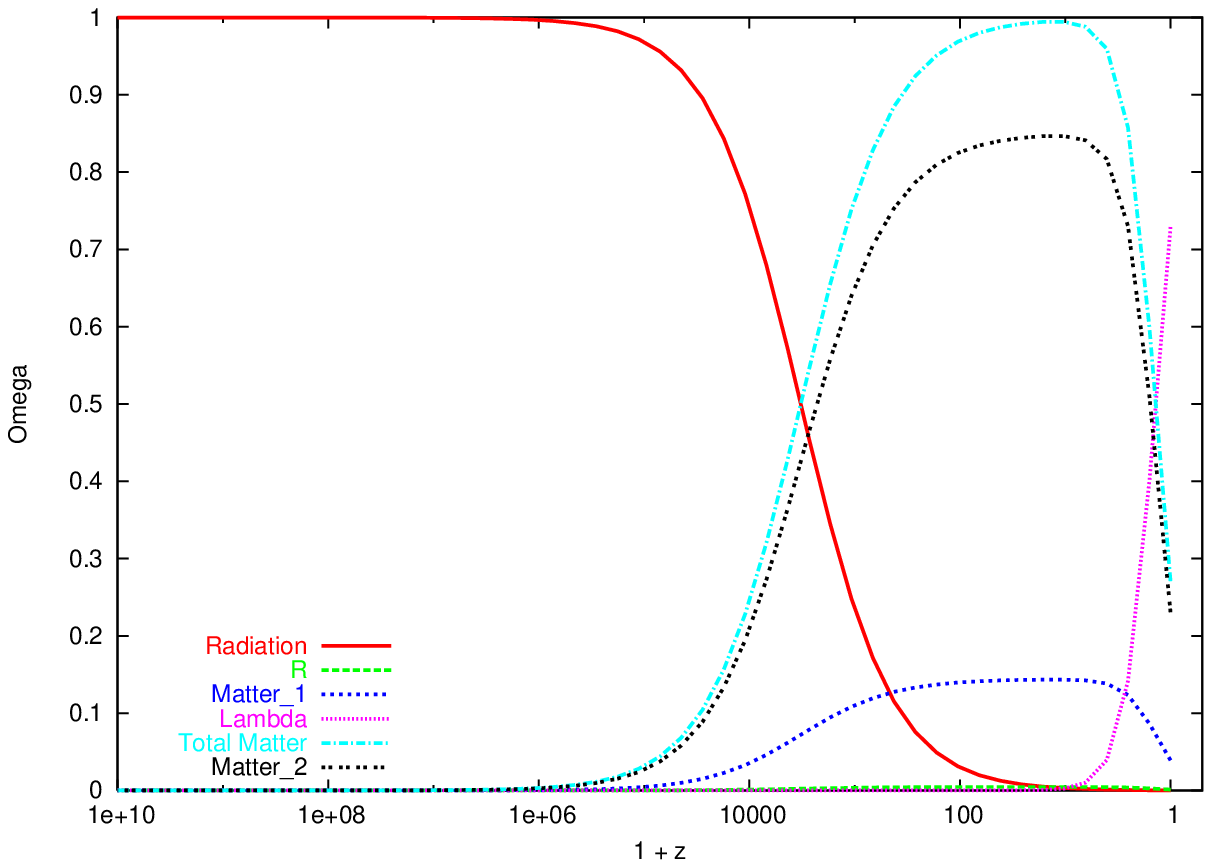}\includegraphics{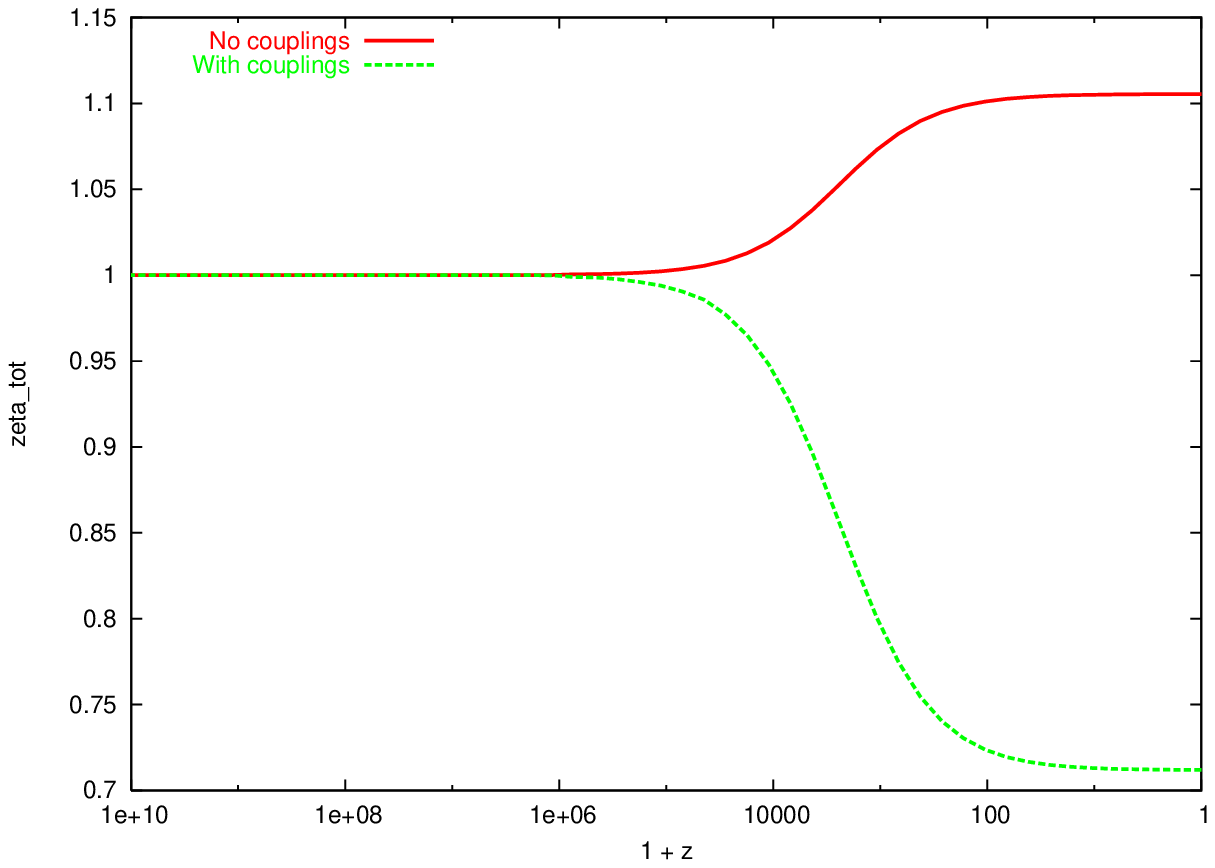}}}
\centerline{\scalebox{0.7}{\includegraphics{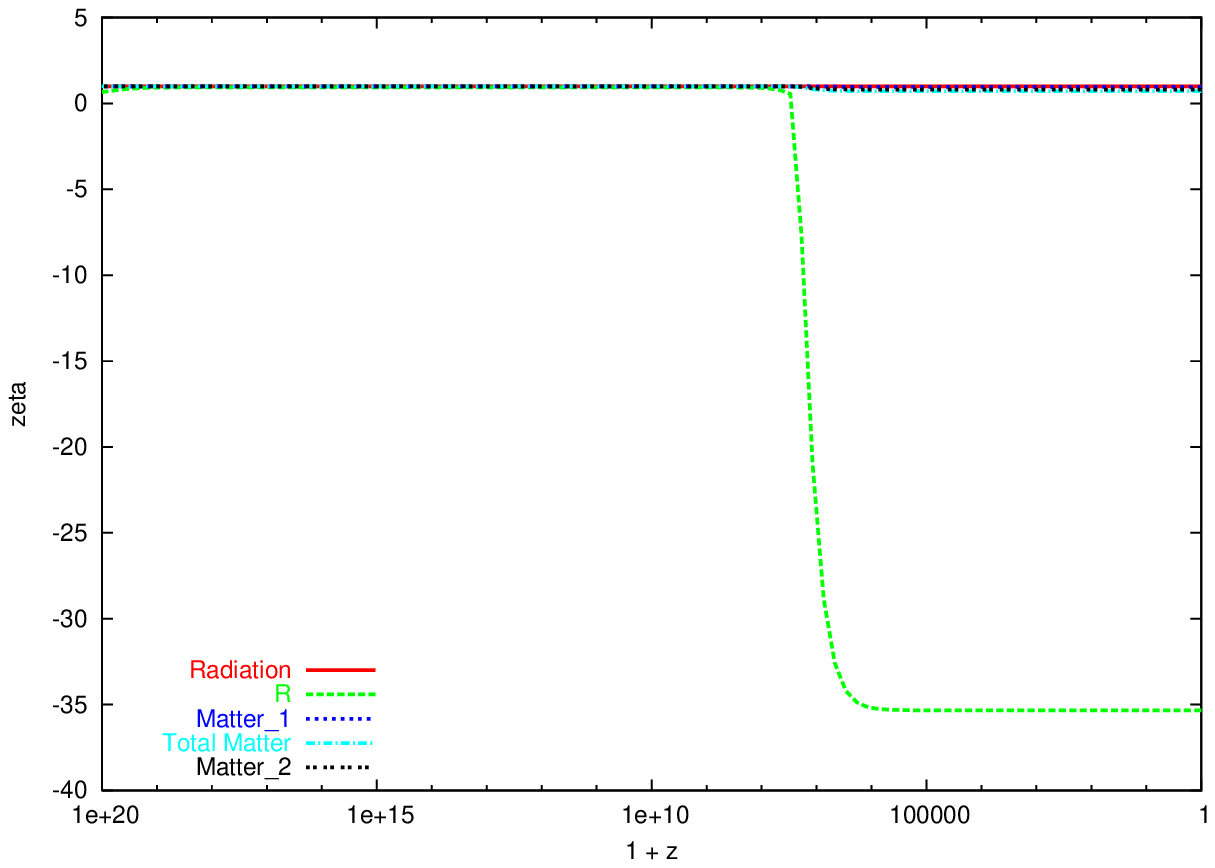}\includegraphics{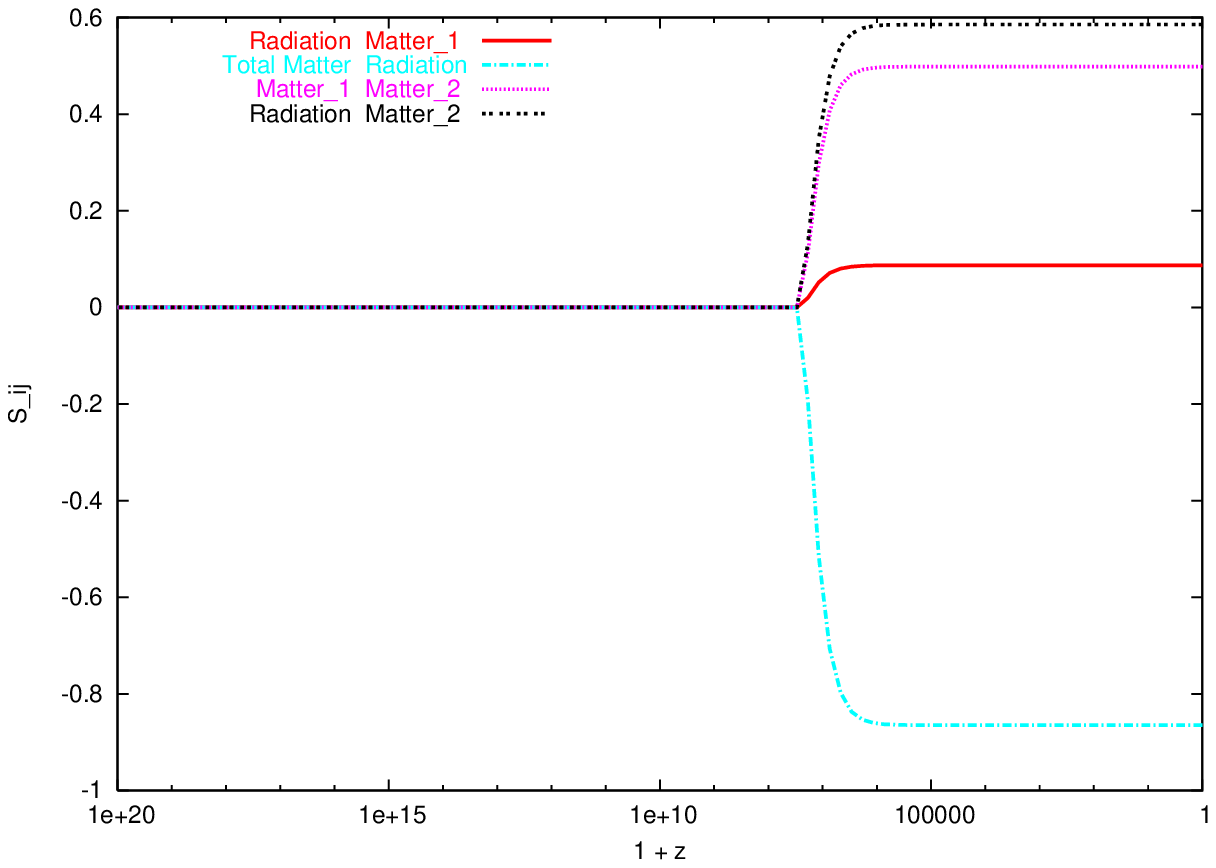}}}
\caption{In this example, we generate a universe with $\Omega_\Lambda = 0.73$,
  $\Omega_R = 0.0001$, $\Omega_{m}^{(2)} = 0.23$, $\Omega_{m}^{(1)} =
  0.04$ today. We have set $M_R = \sqrt{6}\times 10^{-50}\,M_{pl}$.
  The difference between this and the previous Figure \ref{fig:msa2} is that the 
  radion field is subdominant throughout the history of 
  the universe. This is clear from the top left plot in which we show the evolution of the density parameters, $\Omega_i$.
  In the top right plot we show the 
  evolution of the total curvature perturbation, $\zeta$, with and without coupling terms. We set $R_0
  = 1.002\,M_{pl}$ and $R_c = 1.0\,M_{pl}$ so that $\alpha_R^{(1)}\,M_{pl} \sim 0.15$ and
  $\alpha_R^{(2)}\,M_{pl} \sim 1.05$. As one can see, the evolution of the curvature perturbation is 
  significantly modified, even when the radion field is subdominant. The
  lower left plot shows the behavior of the individual $\zeta_i$
  and on the right we show the evolution of the entropy perturbations, $S_{ij}$.}\label{fig:msa1}
\end{figure}

Let us now consider a subdominant radion and examine the entropy
production from the couplings. The results for this setup are shown in Figure~\ref{fig:msa1} where we 
 stabilize the radion at $R_c = 1 \, M_{pl}$.
The couplings can have important consequences for the 
background evolution if the radion is very light. These couplings  dictate 
the evolution of the radion initially and push the radion up its potential. 
This can lead to a period of radion domination as it gains energy from matter. 
We find from our numerical results that if $R$ is to remain subdominant we require 
$M_R \gtrsim 10^{-52} \,M_{pl}$. Once more we generate entropy perturbations.

We have seen that for a long-lived radion its stabilization has some
interesting consequences for the history of the universe. Even if it remains
subdominant in the background evolution, it is still possible for it to generate 
entropy perturbations between matter and radiation due to the couplings. 

\subsubsection{The case of a short-lived radion}
We move on to discuss the case of the decaying radion. In order for
this to occur before nucleosynthesis-- so that the universe is
radiation dominated-- we require $M_R > 10^{-14} \,M_{pl}$. This should
give us a large period of radion domination which is followed by radion decay into 
radiation. In the case of the Goldberger-Wise stabilization mechanism 
via a bulk scalar field, the mass of the radion is of order $1\,{\rm TeV}$ and 
therefore decays naturally in the early universe. In some sense, it could be 
a candidate for the curvaton. Here, however, we are concerned how initial 
perturbations from inflation are modified by the stabilization and decay of the 
radion. We will set the mass of order $100\,{\rm TeV}$ so that the decay occurs 
before nucleosynthesis. 

Again, the background initial conditions are set to
produce the universe that we observe today. As we have already
discussed, $\zeta$ should evolve when $R$ becomes dominant and then
again at matter-radiation equality. If $\alpha^{(j)}_R = 0$, then
$\zeta$ should return to unity. Any deviation from this is down to the
couplings to matter. This can be seen from equation (\ref{eq:zetarho}) 
because $\Omega_R = 0$ and $\zeta_m^{(j)}$ are constant. 

\begin{figure}[!ht]
\centerline{\scalebox{0.7}{\includegraphics{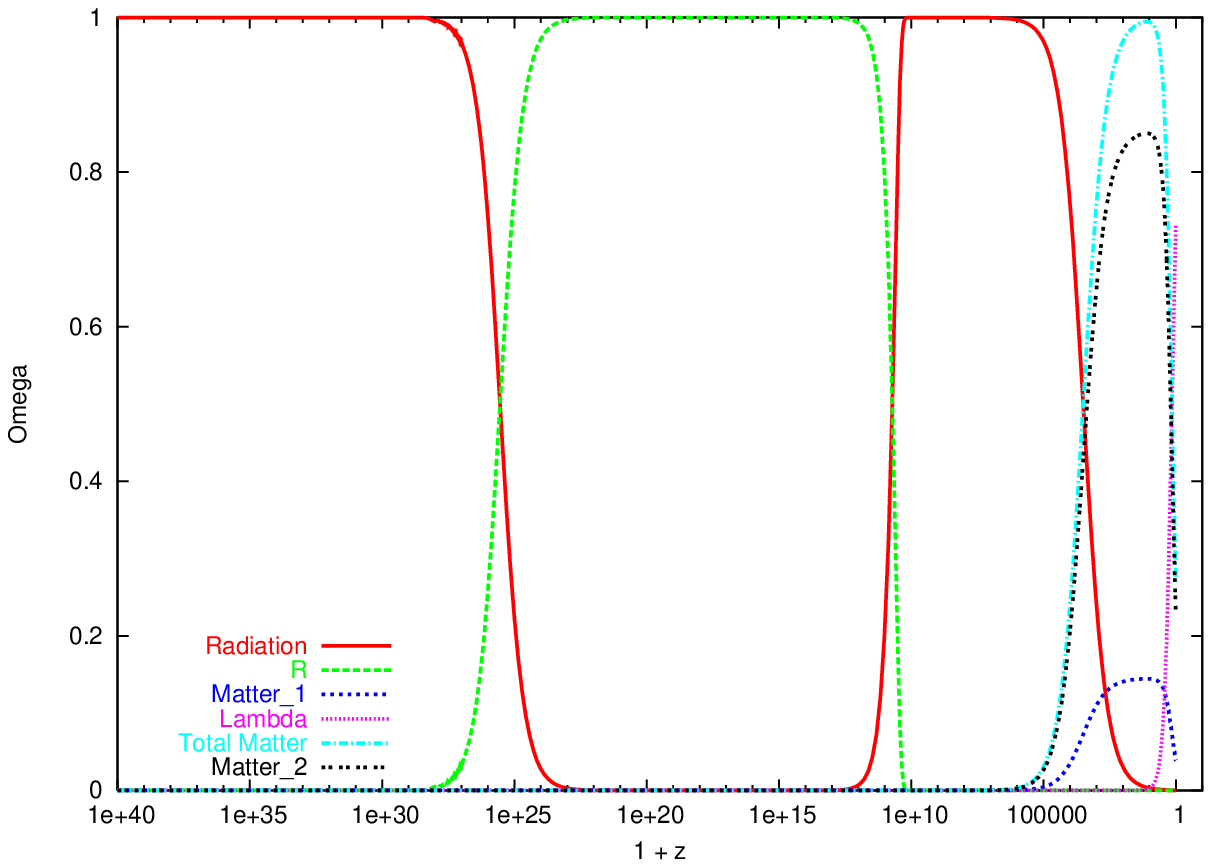}\includegraphics{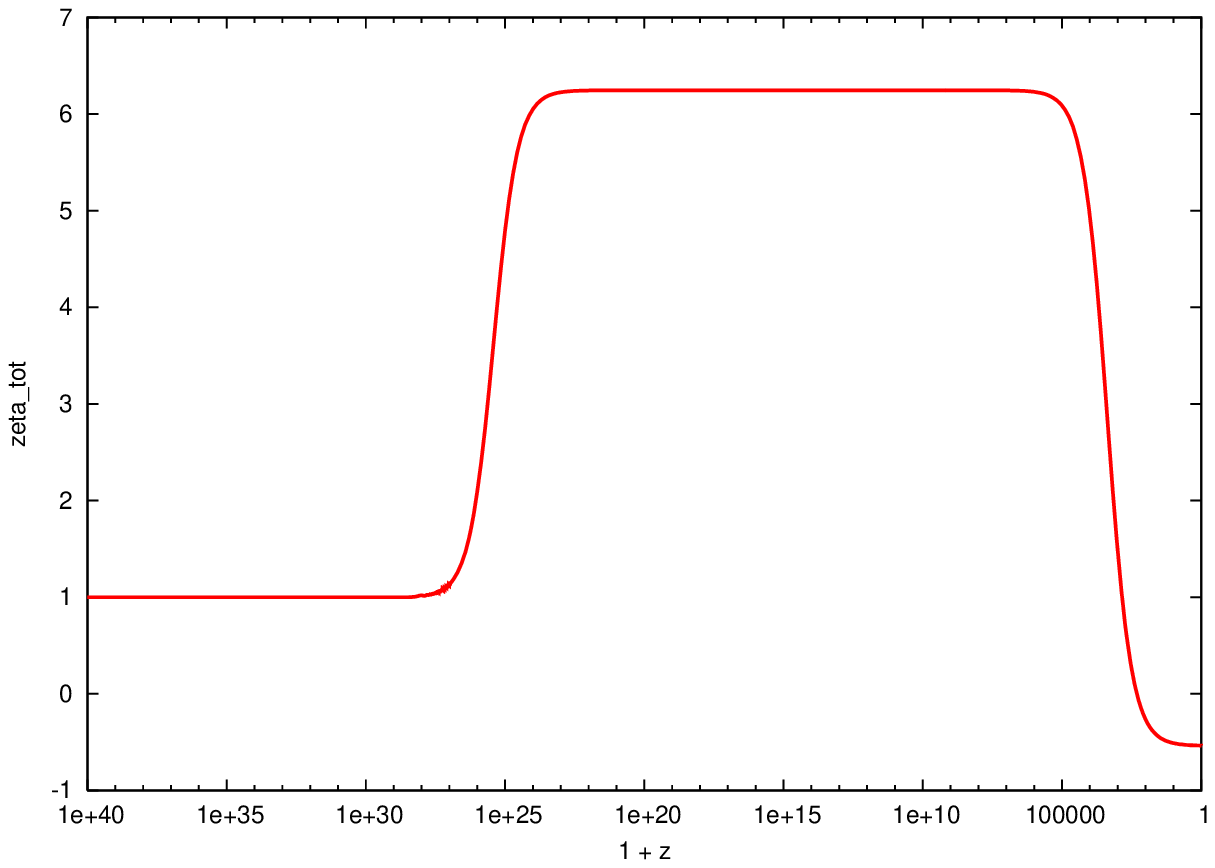}}}
\centerline{\scalebox{0.7}{\includegraphics{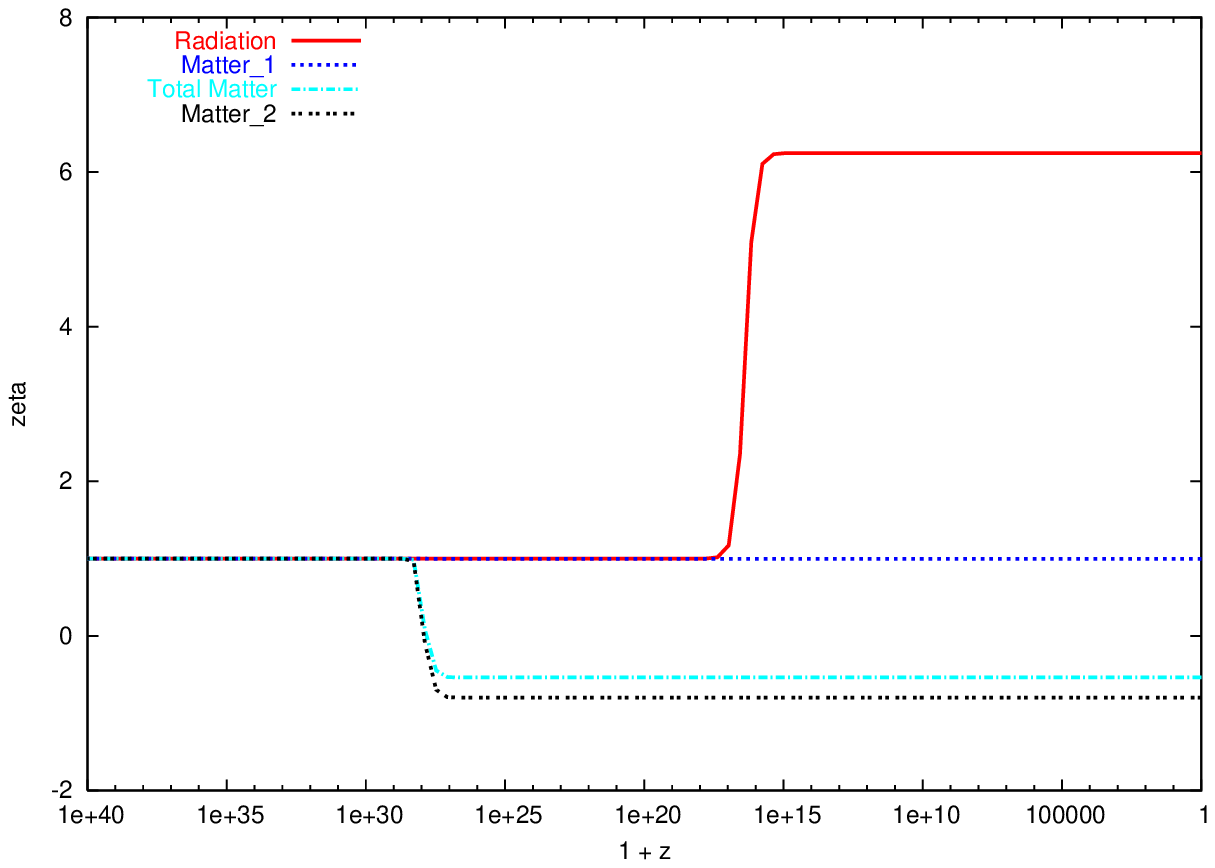}\includegraphics{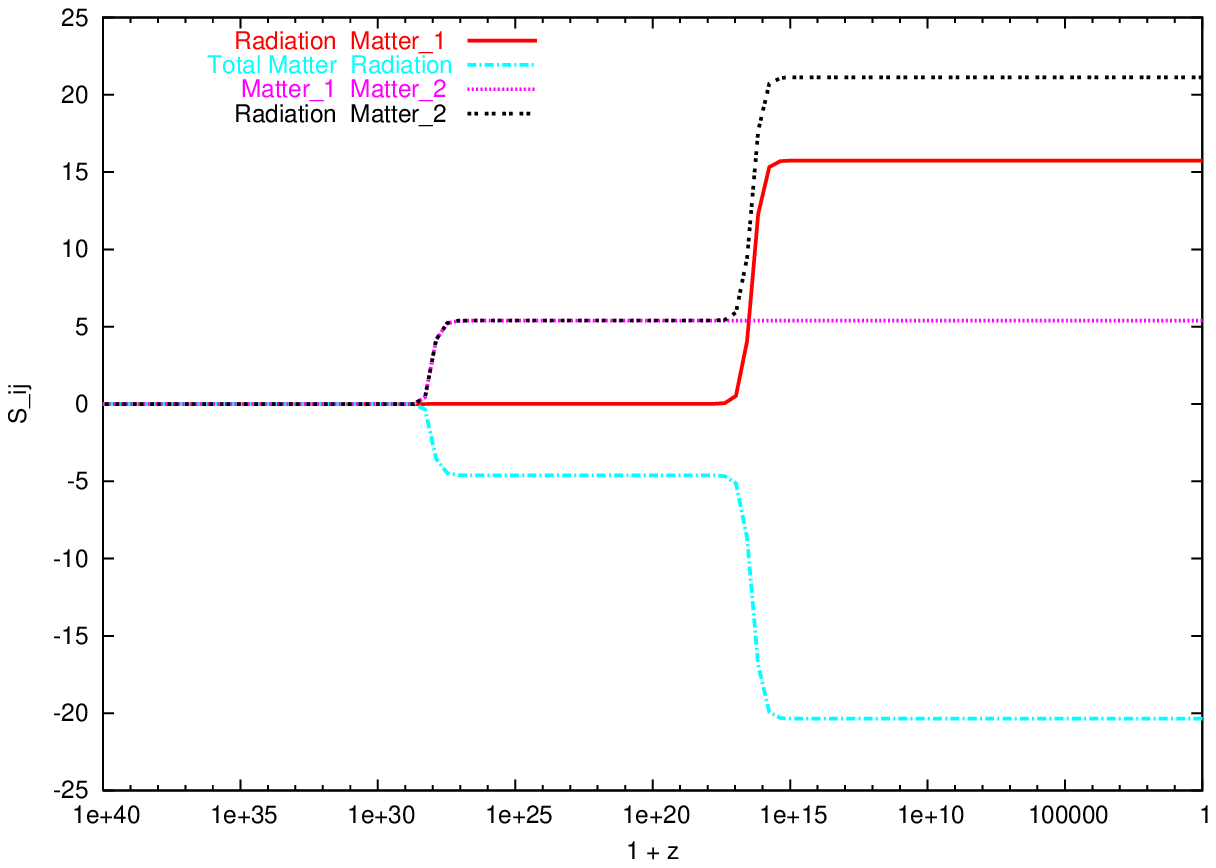}}}
\caption{In this instance, we consider a short-lived radion field with a mass $M_R = \sqrt{6}\times 10^{-13} M_{pl}$.
  We set $R_0 = 0.10\,M_{pl}$ and $R_c = 0.03\,M_{pl}$ so that $10^{-3} \lesssim \alpha_R^{(1)}\,M_{pl} \lesssim 10^{-2}$ and
  $10 \lesssim \alpha_R^{(2)} \,M_{pl}\lesssim 10^2$. This results in a universe with $\Omega_\Lambda = 0.73$,
  $\Omega_{m}^{(2)} = 0.23$, $\Omega_{m}^{(1)} = 0.04$.  In the top left we show the time dependence of 
  the $\Omega_i$. We see that after a period of radion
  domination it decays into radiation. In the top right plot we see that
  evolution of the total curvature perturbation, $\zeta$.
  The lower left panel shows the behavior of the individual $\zeta_i$
  and on the right we show the evolution of all of the entropy
  perturbations, $S_{ij}$, between the different species.}\label{fig:msa2g}
\end{figure}

In Figure~\ref{fig:msa2g} we stabilize the radion close to the horizon of the AdS spacetime  
at $R = 0$ to generate large couplings on
the negative tension brane. One can see that the background dynamics
are as expected and that the radion decays quickly into matter before
nucleosynthesis-- this is ensured by setting $M_R = \sqrt{6}\times 10^{-13}\,
M_{pl}$. The $\zeta_m^{(j)}$ are sourced when the radion is stabilized
and then remain constant, whereas $\zeta_\gamma$ is sourced during the
decay of $R$. This is reflected in the not insignificant production of
entropy. 

\begin{figure}[!ht]
\centerline{\scalebox{0.7}{\includegraphics{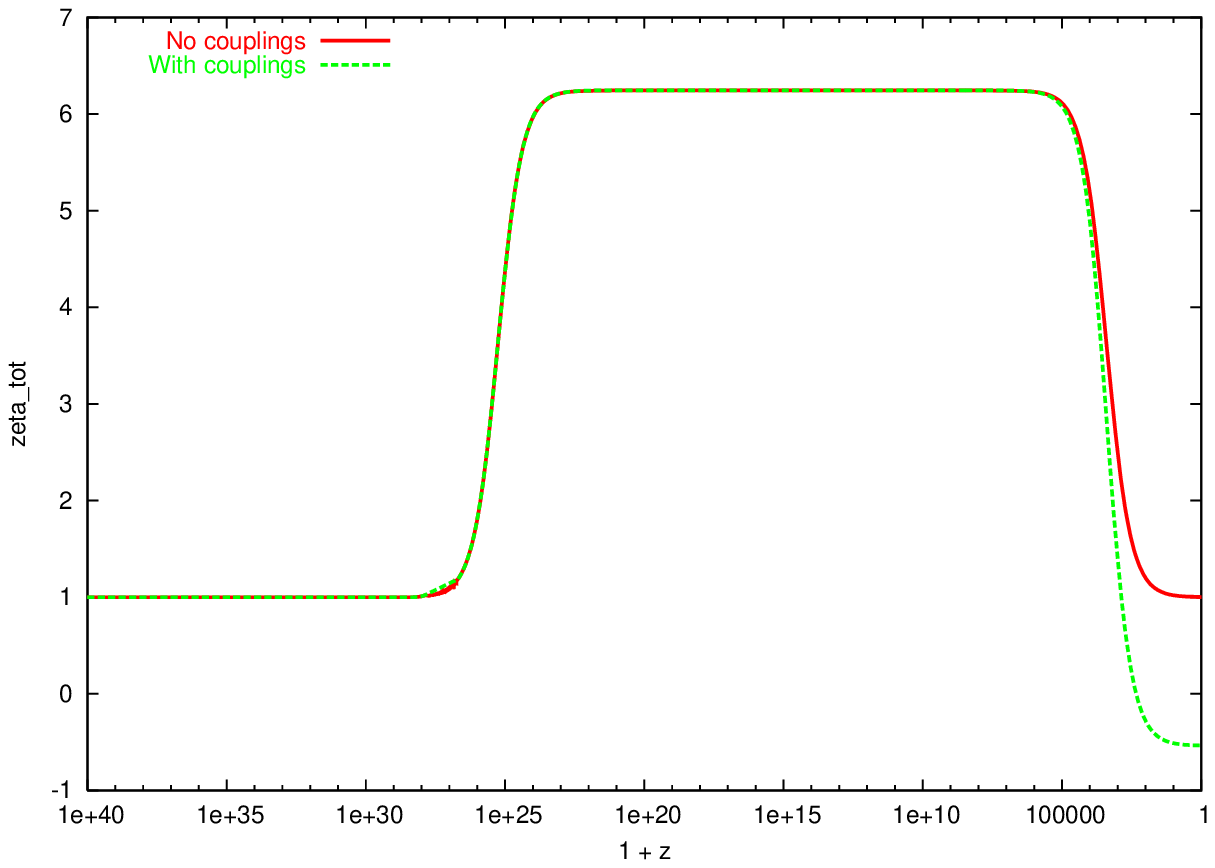}}}
\centerline{\scalebox{0.7}{\includegraphics{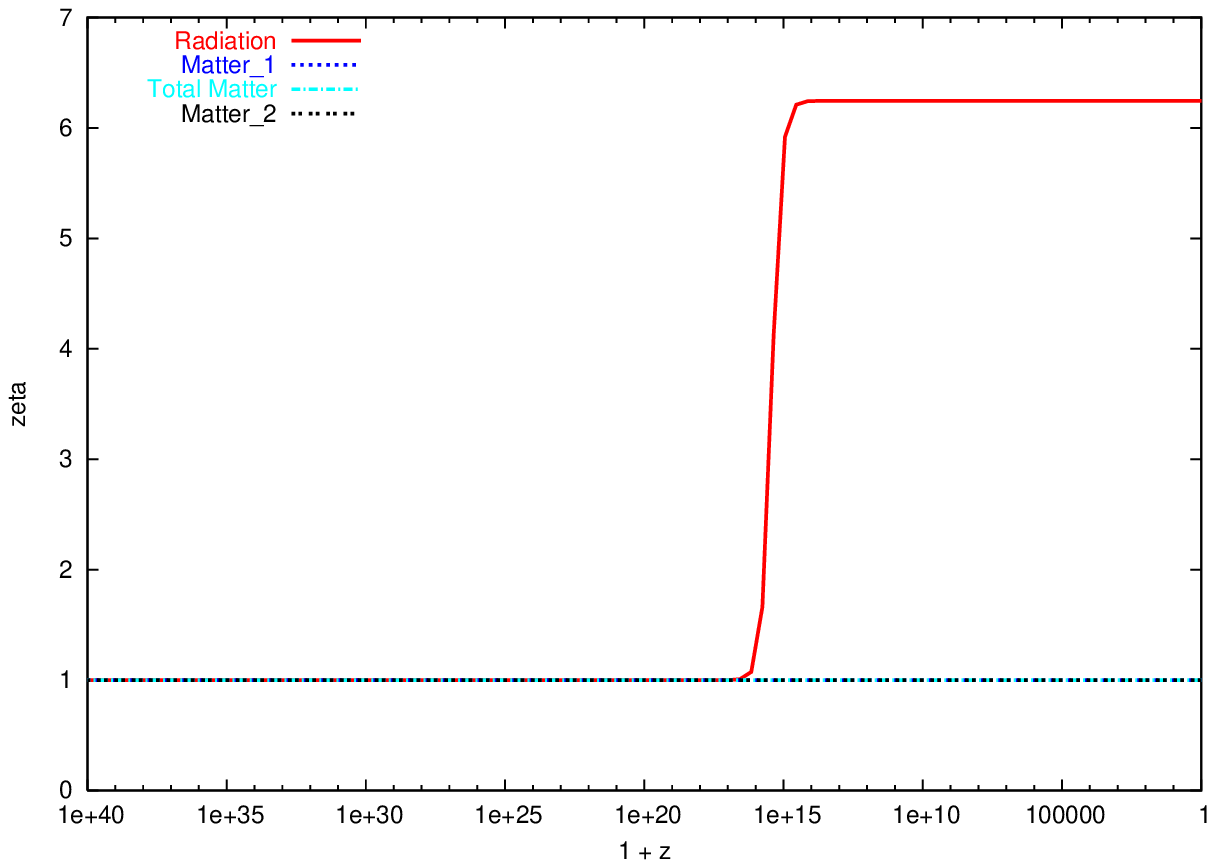}\includegraphics{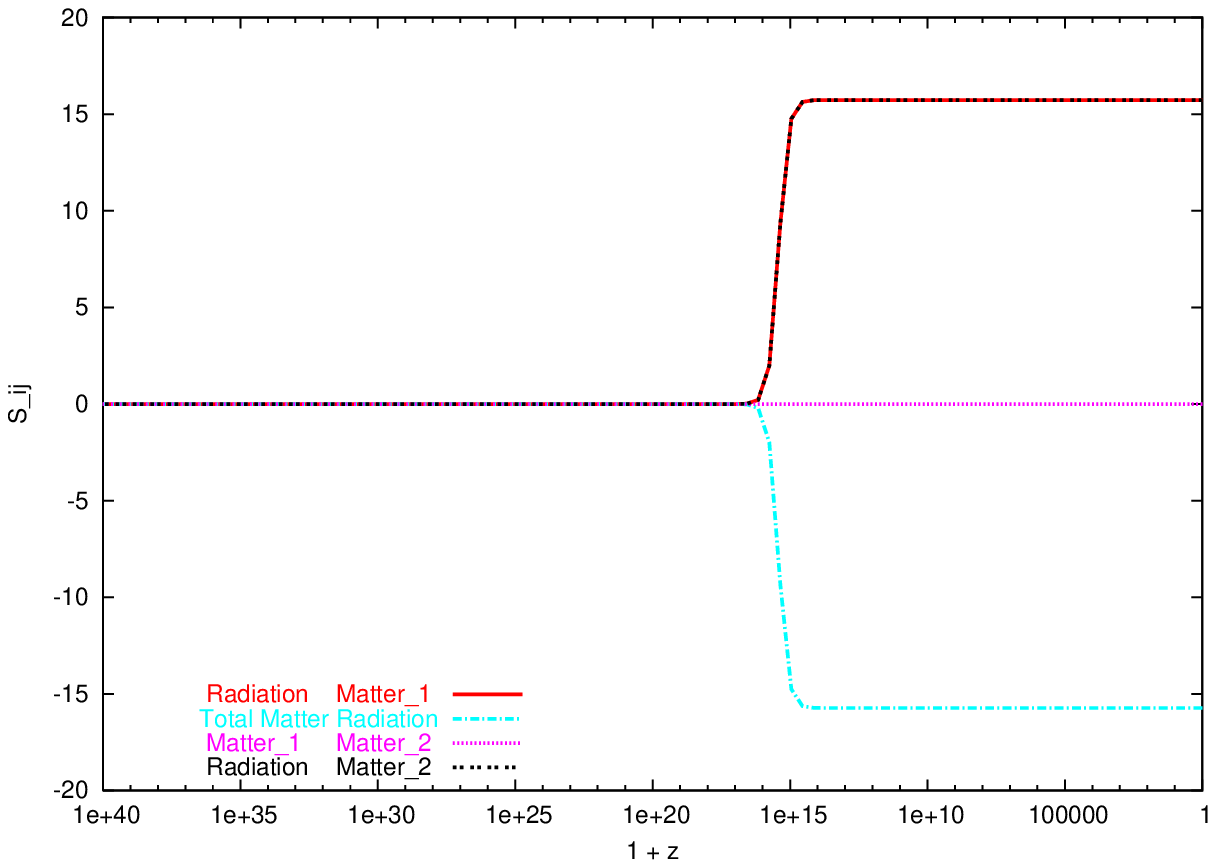}}}
\caption{We make the equivalent run to Figure~\ref{fig:msa2g} but
  without coupling to matter. In the upper plot, we see the
  evolution of the total curvature perturbation, $\zeta$, with no couplings and the equivalent from Figure~\ref{fig:msa2g}. 
  The effects of the couplings on $\zeta$ play a r\^ole only well after radion decay. However, if one compares the 
  two lower left plots, it can be seen that the individual $\zeta_m^{(j)}$ evolve at radion stabilization. With no 
  couplings, it is only $\zeta_\gamma$ that undergoes any change. 
  In the lower right panel, we show the evolution of the entropy perturbations.}\label{fig:nocoupg}
\end{figure}

By means of comparison, we show the equivalent plots in
Figure~\ref{fig:nocoupg} for no coupling to
matter. Whilst the picture is similar, we find $\zeta$ returning to
unity at matter-radiation equality. Furthermore, there is less
entropy production than before.

\section{Summary and Discussion}\label{sect:radconc}
In this paper we have studied the evolution of cosmological perturbations during radion stabilization. 
We have seen that stabilizing the radion field, which describes the inter-brane distance, can have 
non-trivial effects on the curvature and entropy perturbations. 
We have studied two situations, a long and short-lived radion. In the short-lived case, 
we allow the radion to decay into radiation only with the radion coupling to matter naturally.
In setting the initial conditions for the perturbations, we assume that our universe has already undergone a
period of inflation. The scale of inflation determines the initial conditions for the radion-- in particular 
the amplitude of the field fluctuation-- with the fluid perturbations set to be adiabatic
initially. This will naturally source the total curvature perturbation, since there is an initial 
entropy between the fields and the fluids. Additionally, energy is transfered between the radion field and 
matter on each brane due to the coupling terms. This enhances the amount of entropy produced. The amount 
of entropy production coming from the coupling terms depends on the initial conditions of the radion 
field just after inflation as well as on the field value at stabilization. 

The evolution of $\zeta$ and the $\zeta_i$-- and therefore $S_{ij}$-- are 
influenced by a number of parameters. The values 
$R_0$ and $R_c$ are important as they determine the size of the coupling terms $\alpha_R^{(j)}$. 
In addition, $|\delta R_{\rm ini}|$ and $C$ determine the initial entropy perturbation between radiation 
and the field which is transfered into part of the curvature perturbation. This is similar to the 
curvaton scenario. The difference here is that there is an initial curvature perturbation and even 
if the field decays into radiation only, there is still an energy transfer to matter via the couplings.

The constraints on the theory come mainly from the CMB anisotropies. 
The effect of the radion on the CMB anisotropies has been investigated earlier, \cite{cmb1,cmb2}, 
but in this paper we have included an initial field perturbation, $\delta R \neq 0$, set by the inflationary scale.
The large-scale temperature
fluctuation can be written as, \cite{gordonlewis},
\begin{equation}
\frac{\delta T}{T} \approx \frac{1}{5}\zeta_\gamma 
+ \frac{2}{5}\left( \frac{\rho_m^{(1)}}{\rho_M} {\cal S}_{m\gamma}^{(1)} 
+ \frac{\rho_m^{(2)}}{\rho_M} {\cal S}_{m\gamma}^{(2)} + \frac{\rho_R}{\rho_M} {\cal S}_{R\gamma}\right),
\end{equation} 
since the radion behaves like matter at matter-radiation equality in all cases we have considered.
In this expression we have set $\rho_M = \rho_m^{(1)} + \rho_m^{(2)} + \rho_R$ and we neglect any
neutrino contribution. 
Observationally constrained is the ``effective baryon isocurvature perturbation'', defined as 
\begin{equation}
{\cal S}_{\rm eff} = S_{m\gamma}^{(1)} + \frac{\rho_m^{(2)}{\cal S}_{m\gamma}^{(2)}
+\rho_R{\cal S}_{R\gamma}}{\rho_m^{(1)}},
\end{equation} 
where in this expression we have assumed that the matter on the first brane is comprised of baryons 
only. Since the amplitude of ${\cal S}_{m\gamma}^{(2)}$ as well as of ${\cal S}_{R\gamma}$ depend on 
the details of the stabilization mechanism, such as sizes of the couplings during radion evolution 
and the mass of the radion, the CMB constrains these parameters to a certain extent, in a similar 
manner to the curvaton scenario, \cite{gordonlewis}. 

It is possible to extend our work in several different ways. Firstly, the initial conditions between
the matter and radiation do not have to be necessarily adiabatic. In fact, the radion is a candidate 
for a curvaton field, if inflation happens at a higher energy scale than the energy scale of the 
radion potential. In this case, the radion field picks up perturbations during inflation which can
be transfered to a curvature perturbation during radion stabilization. Since in the Goldberger-Wise
mechanism the mass of the radion is of order TeV, this is a realistic scenario to be investigated. 
A detailed analysis would give constraints on the mass of the radion as well as on initial conditions 
of the field just after inflation. We note, that our setup is similar to the one discussed 
in \cite{curvatonappli1}. However, in our case the radion field decays
into radiation only and is   
coupled to matter via the functions $\alpha_R^{(i)}$ only. 

A second point to be investigated is the effect of the potential on the stabilization and on the 
evolution of perturbations. If the radion undergoes minimal evolution then any potential with a 
local minimum can usually be approximated by the expansion
\begin{equation}
V(R) \approx V_0 + \frac{1}{2} M_R^2 (R - R_c)^2 + \dots \label{eq:potapprox}
\end{equation}
In this instance our results will should not change substantially. However, if the radion evolves 
sufficiently such that the potential can no longer be approximated by the expansion in equation~(\ref{eq:potapprox}),
then the energy transfer between the radion and matter can be significantly different. 

Our work also has implications for other models such as those with constant couplings. If these couplings 
are large enough, the energy transfer between the modulus and matter 
can be large and an entropy perturbation will be produced, even if the field is subdominant throughout 
the history of the universe. 

\vspace{0.5cm}
\noindent{\bf Acknowledgements:} The authors are supported by PPARC.

\end{document}